\documentclass[10pt,prd,twocolumn,
nofootinbib,noshowkeys,noshowpacs,
superscriptaddress,floatfix]{revtex4-2}
\usepackage{slashed}
\usepackage{caption}
\usepackage{subfigure}
\usepackage{amsmath,amssymb,graphicx,xcolor,mathtools}
\usepackage[colorlinks=true,linktocpage=green,linkcolor=blue,citecolor=blue]{hyperref}
\allowdisplaybreaks
\usepackage{hyperref}
\hypersetup{
colorlinks=true,
linkcolor=blue,
filecolor=magenta,      
urlcolor=blue,
pdftitle={Shear-Kubo-Rotation}
}
\usepackage{mathrsfs}
\usepackage{simpler-wick}

\DeclareMathOperator{\Tr}{Tr}
\def\be {\begin{equation}}
	\def\ee {\end{equation}}
\def\bea {\begin{eqnarray}}
	\def\eea {\end{eqnarray}}
\def\bc {\begin{center}}
	\def\ec {\end{center}}
\def\nn {\nonumber}

\def\sumintof{\sum\!\!\!\!\!\!\!\!\!\int\limits_{\{P\}}}

\def\wtg{\widetilde{\gamma}}
\def\wtD{\widetilde{D}}

\def\ovpsi{\overline{\psi}}

\DeclareMathAlphabet{\mathpzc}{OT1}{pzc}{m}{it}

\begin{document}
	\preprint{}

\title{Shear viscosity of rotating, hot, and dense spin-half fermionic systems from quantum field theory}
	
	\author{Sarthak Satapathy}
	\email{sarthaks680@gmail.com}
	\affiliation{Department of Physics, Indian Institute of Technology Roorkee, Roorkee - 247667, India}

	\author{Rajeev Singh}
	\email{rajeevofficial24@gmail.com}
    \affiliation{Center for Nuclear Theory, Department of Physics and Astronomy, Stony Brook University, Stony Brook, New York, 11794-3800, USA}
    \affiliation{Department of Modern Physics, University of Science and Technology of China, Hefei, Anhui 230026, China}

	\author{Pushpa Panday}
	\email{pandaypushpa147@gmail.com}
	\affiliation{Department of Physics, Indian Institute of Technology Roorkee, Roorkee - 247667, India}

	\author{Salman Ahamad Khan}
	\email{salmankhan.dx786@gmail.com}
	\affiliation{Department of Physics, Integral University, Lucknow - 226026, India}

 \author{Debarshi Dey}
	\email{debs.mvm@gmail.com}
	\affiliation{Department of Physics, Indian Institute of Technology Roorkee, Roorkee - 247667, India}

\date{\today} 
\bigskip
\begin{abstract}
In this study, we calculate the shear viscosity for rotating fermions with spin-half under conditions of high temperature and density. We employ the Kubo formalism, rooted in finite-temperature quantum field theory, to compute the field correlation functions essential for this evaluation. The one-loop diagram pertinent to shear viscosity is analyzed within the context of curved space, utilizing tetrad formalism as an effective approach in cylindrical coordinates. Our findings focus on extremely high angular velocities, ranging from 0.1 to 1 GeV, which align with experimental expectations. Furthermore, we explore the inter-relationship between the chemical potential and angular velocity within the scope of this study.
\end{abstract}
\maketitle
\section{Introduction}
\label{INTRO}
The study of Quark Gluon Plasma (QGP), formed during heavy-ion collisions and particularly in high electromagnetic fields or significant angular velocities, is a critical area of research~\cite{Bass:1998vz,Kharzeev:2000ph,Heinz:2001xi,Shuryak:2004cy,Gyulassy:2004zy,Florkowski:1321594,Jacak:2012dx,Kharzeev:2013jha,Hattori:2016emy,Hidaka:2022dmn}. High angular velocities in off-center collisions, leading to angular momentum of $10^{3\sim 5}\hbar$, have been observed by the STAR collaboration, adding complexity to the study of these extreme matter states~\cite{Becattini:2021lfq}.
Advances from RHIC and LHC experiments show that QGP rapidly evolves from a glasma state to equilibrium, and then into hadrons~\cite{Kharzeev:2000ph,Heinz:2001xi,Gyulassy:2004zy,Shuryak:2004cy}. Theoretically, relativistic hydrodynamics supports the view of QGP as the smallest fluid droplets with almost perfect fluidity, leading to new explorations in off-equilibrium hydrodynamics~\cite{Romatschke:2007mq, Heinz:2013th,Son:2006em,Schafer:2009dj,Schenke:2021mxx,Florkowski:2017olj}.

Furthermore, measurements by the STAR collaboration indicate that emitted particles, such as $\Lambda$ hyperons, show spin polarization, highlighting QGP as the most vortical system observed in heavy-ion collisions to date~\cite{STAR:2017ckg,Adam:2018ivw,STAR:2019erd,ALICE:2019aid,Acharya:2019ryw,ALICE:2021pzu}. This spin alignment, corresponding with their angular momentum, has intensified interest in the underlying dynamics. Extensive research using various methodologies has been conducted to explore these dynamics~\cite{Son:2009tf,Kharzeev:2010gr,Florkowski:2017ruc,Florkowski:2017dyn,Florkowski:2018ahw,Florkowski:2018fap,Florkowski:2019qdp,Singh:2021man,Florkowski:2021wvk,Bhadury:2020puc,Bhadury:2020cop,Hattori:2019lfp,Fukushima:2020ucl,Li:2020eon,Montenegro:2020paq,Weickgenannt:2020aaf,Garbiso:2020puw,Gallegos:2021bzp,Sheng:2021kfc,Speranza:2020ilk,Becattini:2016gvu,Karpenko:2016jyx,Pang:2016igs,Xie:2017upb,Becattini:2017gcx,Fu:2020oxj,Florkowski:2021xvy,Montenegro:2018bcf,Montenegro:2020paq,Gallegos:2021bzp,Serenone:2021zef,Torrieri:2022xil,Hattori:2019lfp,Fukushima:2020ucl,Li:2020eon,Daher:2022xon,Cao:2022aku,Hu:2021lnx,Hidaka:2018ekt,Yang:2020hri,Wang:2020pej,Weickgenannt:2020aaf,Weickgenannt:2021cuo,Sheng:2021kfc,Hu:2021pwh,Das:2022azr,Weickgenannt:2022zxs,Stephanov:2012ki,Chen:2014cla,Gorbar:2017toh,Hidaka:2018ekt,Shi:2020htn,Heller:2020hnq,Gallegos:2020otk,Garbiso:2020puw,Gallegos:2021bzp,Hongo:2021ona,Gallegos:2022jow,Singh:2022uyy,Singh:2022ltu,Singh:2022pis}. Recent reviews have provided insights into both local and global spin polarization, and developments in spin hydrodynamics, from holographic and quantum field theoretical perspectives, have further enriched this field of study.
Recent trends in the field have expanded to include the exploration of spin polarization within nuclear matter and its interaction with electromagnetic fields. These new measurements have significantly enhanced our comprehension of the Quark Gluon Plasma (QGP), concurrently driving numerous theoretical developments. Consequently, the study of QGP, particularly in extreme conditions, continues to be a dynamic and intriguing area of scientific research.

Theoretical research predicts numerous intriguing phenomena in extreme conditions, like the chiral vortical effect~\cite{Kharzeev:2010gr,Kharzeev:2007tn,Burnier:2011bf,Kharzeev:2015znc}, chiral vortical wave~\cite{Jiang:2015cva}, mass splitting under rotation~\cite{WeiMingHua:2020eee}, magnetic chiral density wave~\cite{Ghalati:2023npr}, and vortical effects in AdS space~\cite{Ambrus:2021eod}. These anomalous processes, reliant on experimental signatures, are observable in heavy-ion collision experiments. High angular velocities in the QCD medium provide an opportunity to explore these properties in detail, facilitating the extraction of experimental signatures.
Another intriguing research area is the QCD phase diagram under rotation~\cite{Becattini:2021lfq,Chernodub:2020qah,Sadooghi:2021upd}. Studies have shown that rotation affects the confining and deconfining phases of QGP. For instance, the deconfining transition occurs at a specific distance from the rotation axis~\cite{Chernodub:2020qah}. There is evidence of a mixed inhomogeneous phase with spatially separated confinement and deconfinement regions, leading to the prediction of two deconfining temperatures. In the context of lattice SU(3) gauge theory, the isothermal moment of inertia for a rigidly rotating QGP has been computed, revealing negative values below the supervortical temperature, which is 1.5 times the critical temperature~\cite{Braguta:2023yjn}. This suggests that a rigidly rotating system becomes thermodynamically unstable beyond a certain temperature.
 
In the study of rotating QCD matter, transport coefficients are vital for hydrodynamic simulations~\cite{Hu:2021lnx}. These coefficients can be calculated using approaches like kinetic theory, based on the relativistic Boltzmann equation~\cite{Bhadury:2020cop,Bhadury:2020puc}, and field theoretic formulations, which focus on correlation function calculations~\cite{Hongo:2021ona,Harutyunyan:2018wdk,Harutyunyan:2021rmb,Czajka:2018bod,Becattini:2019dxo}. Recent studies using Zubarev's non-equilibrium statistical operator formalism have reported Kubo formulas for various transport coefficients in first-order spin hydrodynamics, introducing new coefficients such as rotational viscosity and boost heat conductivity in addition to traditional coefficients like shear viscosity, bulk viscosity, and thermal conductivity~\cite{Hu:2021lnx}.
These formulations are essential for understanding the impact of rotational effects, which induce background torsion, interacting with fermionic fields. However, gauge fields do not couple with torsion due to the SU(3)$_\text{c}$ gauge invariance~\cite{Hehl:1976kj,Hongo:2021ona}. The influence of rotation in field theory is akin to the effects of a background magnetic field, altering the propagator's translational symmetry between two spacetime points~\cite{Kuznetsov:2004tb,Kuznetsov:2013sea,Chyi:1999fc,Iablokov:2020upc}. For magnetic fields, this alteration manifests as the Schwinger phase factor, an exponential function indicating a phase shift. In a rotating environment, a similar factor appears in the fermion propagator. Interestingly, translational symmetry is restored at high angular velocities, simplifying the computation of field correlation functions~\cite{Ayala:2021osy}.

In our study, we investigate how finite rotation impacts transport coefficients, particularly focusing on shear viscosity in a fermion-dominated medium subject to high angular velocity. Employing the Kubo formalism, we incorporate the influence of rotation in our calculations by using the momentum space propagator for fermions, derived via the Fock-Schwinger method. The role of angular velocity is notably apparent in two areas: it modifies the distribution function through the summation of Matsubara frequencies, and it is also present in the numerator of the shear viscosity expression via trace evaluations. Our initial findings suggest that both temperature and angular velocity lead to an increase in shear viscosity, while an increase in chemical potential results in its reduction. Interestingly, the study uncovers that chemical potential and angular velocity, though functionally similar, exert opposite effects on shear dissipation within fermionic systems.
 
This paper is structured as follows: Section~\ref{F-Rot} explores the dynamics of fermions in a rotating environment. Next, in Section~\ref{spec}, we delve into the calculations of the spectral function and the shear viscosity of fermions. Section~\ref{Res} is devoted to discussing the results of our study. Finally, we provide a summary in Section~\ref{sec:summary}.
\section{Fermions in a rotating medium}
\label{F-Rot}
In this section, we provide a concise introduction to the behavior of fermions in a rotating setting. To investigate this complex system, our approach will be based in curved spacetime. The metric tensor, akin to that of a curved spacetime, serves as a useful tool for characterizing the geometric properties of the region generated following non-central collisions, which rotates at an angular velocity $\Omega$ about the $z$-axis. For this specific system, the metric tensor $g^{\mu\nu}$ is represented by 
	\bea
	g_{\mu\nu} = \begin{pmatrix}
		1 - \left(x^2 + y^2\right)\Omega^2 & y\Omega & -x\Omega & 0\\
		y\Omega  &  -1 & 0 & 0  \\
		-x\Omega & 0 & -1 & 0 \\
		0 & 0 & 0 & -1 
	\end{pmatrix},
	\label{F-Rot-1}
	\eea 
and the Dirac equation describing a massive fermion with mass $m$ and spin-$\frac{1}{2}$ in cylindrical coordinates~\cite{Ayala:2021osy,Fang:2021mou} is 
\bea
\left(i\wtg^\mu\widetilde{D}_\mu - m \right)\psi = 0\,,
	\label{F-Rot-2}
	\eea  
where $\wtg^\mu$ are the gamma matrices and $\widetilde{D}_\mu$ is the covariant derivative
\begin{equation}
 \widetilde{D}_\mu = \partial_\mu + \Gamma_\mu\,.
 \end{equation}
Here, $\Gamma_\mu = (1/8) \,\omega_{\mu a b}\big[\gamma^a , \gamma^b\big]$ is the affine connection with $\omega_{\mu a b}$ being the spin connection. To compute $\omega_{\mu a  b}$ one has to use the definition of vierbein (also known as tetrad) and metric tensor $g_{\mu\nu}$ which are given as 
	\bea
	e_a^{~\mu} &=& \delta_a^{~\mu} - \delta_a^{~0}\delta_i^{~\mu}v_i\,, \quad \text{with} \quad (i = 1,2,3) \nn\\
    e^a_{~\mu} &=& \delta^a_{~\mu} + \delta^a_{~i}\delta^0_{~\mu}v_i \,,\nn\\
    g_{\mu\nu} &=& \eta_{ab}e^a_{~\mu}e^b_{~\nu}\,,
	\label{F-Rot-3}
	\eea  
where $\eta_{ab} = \text{diag}(1,-1,-1,-1)$ is the Minkowski spacetime metric tensor. Thus, one obtains
	\bea 
	\omega_{\mu a b} = g_{\alpha\beta}e_a^{~\alpha}\left[\partial_\mu e_b^{~\beta} + \Gamma^\beta_{~\mu\nu}e_b^{~\nu}  \right],
 \label{F-Rot-4}
 \eea
where $\Gamma^\beta_{~\mu\nu}$ is the Christoffel symbol defined as~\cite{Singh:2020rht}
 \bea
 \Gamma^\beta_{~\mu\nu} = \frac{g^{\beta\alpha}}{2}\left[ \partial_\nu g_{\alpha\mu} + \partial_\mu g_{\alpha\nu}- \partial_\alpha g_{\mu\nu}  \right].
	\eea 
In curved spacetime the energy-momentum tensor is given as~\cite{Fang:2021mou}
	\bea
	T^{\mu\nu} =  \frac{i}{4}\left(\overline{\psi}\wtg^\mu \wtD^\nu\psi + \overline{\psi}\wtg^\nu \wtD^\mu\psi  \right) + \rm{h.c},
	\label{F-Rot-5}
	\eea 
	where $\psi$ and $\overline{\psi}$ are the Dirac field operator and its conjugate, respectively, and h.c denotes the Hermitian conjugate. In a uniformly rotating frame, $\widetilde{D}^\mu$ is 
	\bea
	\widetilde{D}_\mu = \left( \partial_t - i\frac{\Omega\Sigma_3}{2}, -\partial_x, - \partial_y, -\partial_z \right),
	\label{F-Rot-6}
	\eea 
	where $\Sigma_3 = (i/2)\left[\gamma^1, \gamma^2\right]$. The spacetime dependent gamma matrices $\wtg^\mu$~\cite{Fang:2021mou,Ayala:2021osy} in tetrad system present in Eq.~(\ref{F-Rot-5}) are defined as
	\bea
&&\wtg^0 = \gamma^0\,, \quad  \quad  \quad ~ \quad \wtg_0 = \gamma^0 - x\Omega\gamma^2 + y\Omega\gamma^1\,,\nn\\ 
&&\wtg^1 = \gamma^1 + y\Omega\gamma^0\,, \quad \wtg_1 = -\gamma^1\,,\nn\\
&&\wtg^2 = \gamma^2 - x\Omega\gamma^0\,, \quad \wtg_2 = -\gamma^2\,,\nn\\
&&\wtg^3 = \gamma^3\,, \quad  \quad  \quad ~~ \quad \wtg_3 = -\gamma^3\,,
\label{spec-6}
\eea  
where $\gamma^\mu$ are the gamma matrices in Minkowski space.
In off-central heavy-ion collisions the direction of rotation is perpendicular to the reaction $(x-y)$ plane. Following the above mentioned details, the Lagrangian with finite chemical potential $\mu$ for a medium rotating with constant angular velocity $\Omega$ is expressed as~\cite{Wei:2021dib} 
	\bea
	\mathcal{L} = \overline{\psi}\left[i\gamma^\mu\partial_\mu + \gamma^0\left( \Omega J_z + \mu \right)- m\right]\psi\,,
	\label{F-Rot-7}
	\eea 
	with $J_z$ being the third component of the total angular momentum given by $\vec{J} = \vec{x}\times \vec{p} + \vec{S}$ where $\vec{S} = (1/2)\begin{pmatrix}
		\vec{\sigma} & 0\\
		0 & \vec{\sigma}
	\end{pmatrix}$.
The fermion propagator in the momentum space derived from the above Lagrangian using Fock-Schwinger method is given by~\cite{Iablokov:2020upc,Ayala:2021osy} 
 \begin{widetext}
    \bea
	S(p) &=& \frac{\big( p_0 + \frac{\Omega}{2} - p_z + ip_\perp   \big)\big(\gamma_0 + \gamma_3\big) + m\big(1 + \gamma_5\big)}{\big(p_0 + \frac{\Omega}{2}\big)^2 - \vec{p}^2 - m^2 + i\epsilon}\mathcal{O}^+ + \frac{\big( p_0 - \frac{\Omega}{2} + p_z - i p_\perp   \big)\big(\gamma_0 - \gamma_3\big) + m\big(1 + \gamma_5\big)}{\big(p_0 - \frac{\Omega}{2}\big)^2 - \vec{p}^2 - m^2 + i\epsilon}\mathcal{O}^- \,,
	\label{F-Rot-8}
	\eea 
  \end{widetext}
where $p_0$ is the temporal component, $p_z$ is the $z$ component, parallel to the axis of rotation, and $p_\perp = \sqrt{p_x^2 + p_y^2}$ is the transverse component of the four-momentum, and $\mathcal{O}^{\pm} \equiv (1/2)\left[1 \pm i\gamma^1\gamma^2\right]$.
In the following, we employ Eq.~(\ref{F-Rot-8}) to perform calculations at finite temperature using Imaginary Time Formalism (ITF).
 \begin{figure}[h!]
\centering
\includegraphics[scale = 0.18]{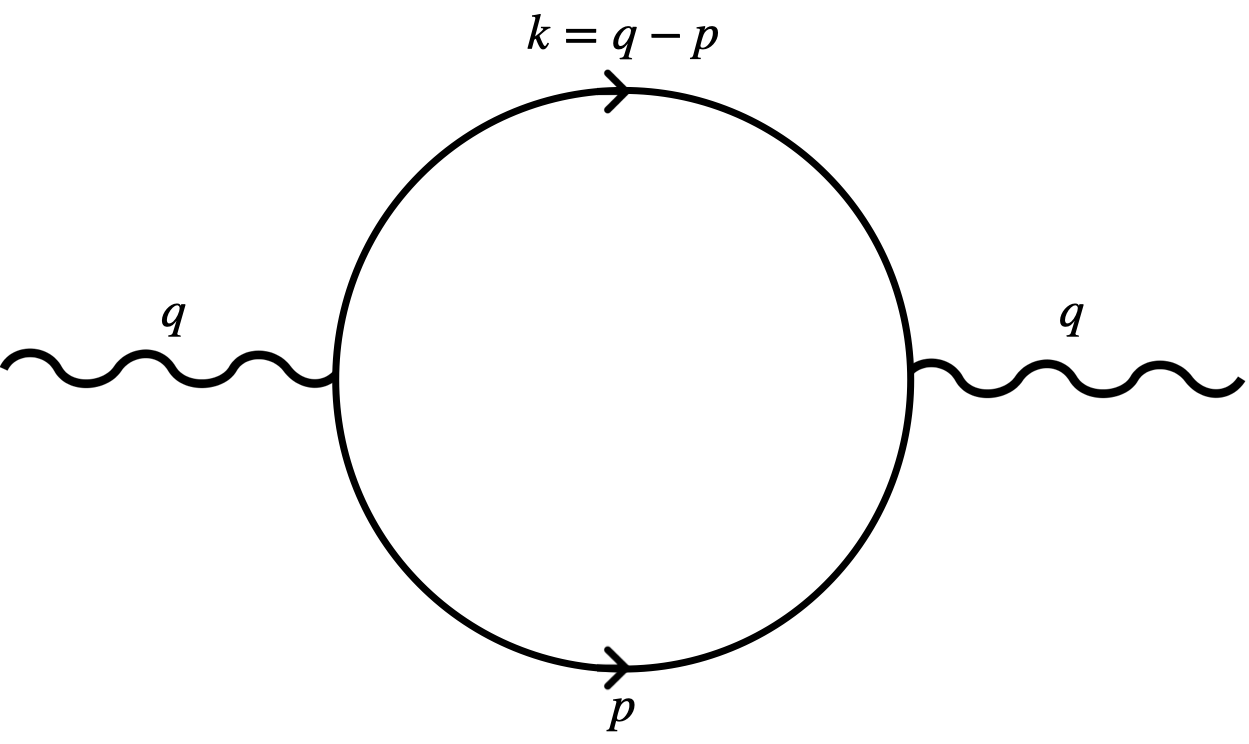}
\caption{One loop diagram representing the photon polarization tensor with $q$ being the momentum of the photon, and $k$ and $p$ being the momentum of the anti-fermion and fermion, respectively.}
\label{fig:oneloop}
\end{figure}
\section{Spectral function and Shear viscosity of fermionic systems from correlation functions}
\label{spec}  
This section demonstrates calculating the shear viscosity of a rotating, hot, and dense fermionic system using the Kubo formalism, which involves computing correlation functions in finite temperature quantum field theory.

We begin with the Kubo formula for shear viscosity
	\bea
	\eta = -\frac{1}{10}\displaystyle{\lim_{q_0 \to 0}}\frac{\rho_\eta(q_0)}{q_0}\,,
	\label{spec-1}
	\eea 
where $\rho_\eta(q)$ is the spectral function of shear viscosity calculated from the two-point correlation function of the shear stress tensor, $\pi^{\mu\nu} =  \Delta^{\mu\nu}_{~~~\alpha\beta}\,T^{\alpha\beta}$, and $q_0$ is the temporal component of the four-momentum. The expression of $\rho_\eta(q)$ is given by 
	\bea
	\rho_\eta(q) = \text{Im}\left[\Pi_\eta(q)\right]\,,
 \label{spec-1A}
 \eea
 where
 \bea
 \Pi_\eta(q) =  i\int d^4r ~e^{iq\cdot r} \big\langle \pi^{\mu\nu}(0)\pi_{\mu\nu}(r)  \big\rangle_R\,.  
 \label{eq:PiEta}
	\eea 
The term $\big\langle\mathcal{O}_1(x)\mathcal{O}_2(y)\big\rangle_R$ represents the retarded thermal average of the composite field expressions $\mathcal{O}_1(x)$ and $\mathcal{O}_2(y)$, positioned at spacetime points $x$ and $y$, respectively. The subscript `R' indicates a retarded two-point function.
 
Equation~(\ref{spec-1A}) is formulated in cylindrical coordinates, with the spacetime coordinate at any point denoted as $r = (t, \rho, \phi, z)$. To advance the calculations, it's necessary to carry out Wick's contraction of fields, as follows
	\bea
	\wick{\c\psi(0)  \c\ovpsi(r)} = S(0,r) = \int\frac{d^4p}{(2\pi)^4}e^{-ip\cdot r}\left(-iS(p)\right)\,,
	\label{spec-7}
	\eea    
where $S(0,r)$ and $S(p)$ is the fermion propagator in coordinate and momentum space, respectively.
Using Eq.~(\ref{spec-7}), the two-point correlation function of energy-momentum tensors at two spacetime points is calculated as 
 \begin{widetext}
	\bea
\langle T^{\mu\nu}(x)T^{\alpha\beta}(y)\rangle 
&=& -\frac{1}{16}\Big\langle\left[\ovpsi\wtg^{\{\mu} (\wtD^{\nu\}}\psi)  - (\wtD^{\{\mu} \ovpsi)\wtg^{\nu\}}\psi \right]_x \left[\ovpsi\wtg^{\{\alpha} (\wtD^{\beta\}}\psi) - (\wtD^{\{\alpha} \ovpsi)\wtg^{\beta\}}\psi \right]_y\Big\rangle \,, \nn \\
&=& -\frac{1}{16}\Big[\big[ \wick{ \c2\ovpsi\wtg^{\{\mu} (\wtD^{\nu\}}\c1\psi) \big]_x\big[  \c1\ovpsi\wtg^{\{\alpha} (\wtD^{\beta\}}\c2\psi)  \big]_y} -  \wick{\big[\c2\ovpsi\wtg^{\{\mu} (\wtD^{\nu\}}\c1\psi)\big]_x\big[(\wtD^{\{\alpha} \c1\ovpsi)\wtg^{\beta\}}\c2\psi\Big]_y }    \nn \\ 
	&& - \wick{\big[(\wtD^{\{\mu} \c2\ovpsi)\wtg^{\nu\}}\c1\psi  \big]_x\big[ \c1\ovpsi\wtg^{\{\alpha} (\wtD^{\beta\}}\c2\psi)   \big]_y} + \wick{\big[  (\wtD^{\{\mu} \c2\ovpsi)\wtg^{\nu\}}\c1\psi  \big]_x\big[  (\wtD^{\{\alpha} \c1\ovpsi)\wtg^{\beta\}}\c2\psi  \big]_y}  \Big]\,,
	\label{spec-8}
	\eea  
 \end{widetext}
where $A^{\{\alpha\beta\}} = A^{\alpha\beta} + A^{\beta\alpha}$.
In Eq.~(\ref{spec-8}) we have considered the general spacetime points $x$ and $y$, which can be specified for cylindrical coordinates at the points $r$ and $r'$, where $r = (t,\rho, \phi, z)$. Here, $t$ is the time, $\rho$ is the cylinder radius, $\phi$ is the azimuthal angle and $z$ is the height of the cylinder. On substituting Eq.~(\ref{spec-7}) in Eq.~(\ref{spec-8}), the general expression for the two-point function of energy-momentum tensors in cylindrical coordinates is 
\begin{widetext}
	\bea
	 \big\langle T^{\mu\nu}(0)T^{\alpha\beta}(r) \big\rangle 
	&=& -\frac{1}{16}\Big[\Tr\big\{ \wtg^{\{\mu}\wtD^{\nu\}} S(0,r)\wtg^{\{\alpha} \wtD^{\beta\}} S(r,0)  \big\} - \Tr\big\{\wtg^{\{\mu} \wtD^{\nu\}}\wtD^{\{\alpha} S(0,r)\wtg^{\beta\}} S(r,0)  \big\} \nn \\
	&& - \Tr\big\{ \wtg^{\{\mu} S(0,r)\wtg^{\{\alpha}\wtD^{\beta\}}\wtD^{\nu\}} S(r,0)   \big\}  + \Tr\big\{ \wtg^{\{\mu}\wtD^{\{\alpha} S(0,r)\wtg^{\beta\}} \wtD^{\nu\}} S(r,0)\big\}   \Big].
	\label{spec-9} 
	\eea 
 \end{widetext}
Ref.~\cite{Ayala:2021osy} demonstrates that under rotation, the propagator's translational invariance is compromised due to the rotation axis favoring a specific direction. A parallel phenomenon occurs in the presence of a background magnetic field~\cite{Kuznetsov:2004tb,Kuznetsov:2013sea,Chyi:1999fc}, where the translational invariance of the propagator is disrupted by a phase shift known as the Schwinger phase factor. In our case, considering a very large $\Omega$ around 0.1 GeV, the phase factor effectively vanishes resulting in the propagator's dependence solely on relative coordinates
$$S(r,r') \xrightarrow{\text{Large~} \Omega} S(r-r').$$
Such a situation can be anticipated in the early stages of off-central heavy-ion collisions where $\Omega$ is very large.
Using Eq.~(\ref{spec-9}) in Eq.~(\ref{eq:PiEta}) and utilizing the translational invariance property of the propagator, $\Pi_\eta(q)$ can be calculated as
\begin{widetext}
\bea
\Pi_\eta(q) &=& i\int d^4r~ e^{iq\cdot r}\big\langle  \Delta^{\mu\nu\alpha\beta}T_{\alpha\beta}   \Delta_{\mu\nu\rho\lambda}T^{\rho\lambda}  \big\rangle_R \nn \\
	&=& \frac{i}{40}\int d^4r\int\int\frac{d^4k}{(2\pi)^4}\frac{d^4p}{(2\pi)^4} e^{i(q-p-k)\cdot r} \Bigg[  (k_z - p_z)^2\Big\{ \Tr\big\{\wtg_1 S(p)\wtg^1 S(k) + \wtg_3 S(p)\wtg^3 S(k)\big\}    \Big\} \nn \\ &&+ \frac{1}{3}\, \Bigg\{   -2 \Tr\Big\{\wtg^0\Big(p_0 + \frac{\Omega\Sigma_3}{2}\Big)S(p)\wtg^0 \Big(k_0 + \frac{\Omega\Sigma_3}{2}\Big) S(k)\Big\}  + \Tr\Big\{ \wtg^0\Big(p_0 + \frac{\Omega\Sigma_3}{2}\Big)^2 S(p) \wtg_0 S(k)\Big\} \nn \\ &&  +  \Tr\Big\{ \wtg^0 S(p) \wtg_0 \Big(k_0 + \frac{\Omega\Sigma_3}{2}\Big)^2S(k)\Big\}  + \Tr\Big\{ \wtg^0\Big(p_0 + \frac{\Omega\Sigma_3}{2}\Big)S(p)\wtg^3 k_z S(k)  \Big\} \nn \\ && + \Tr\Big\{ \wtg^0p_z\Big(p_0 + \frac{\Omega\Sigma_3}{2}\Big)S(p)\wtg_3S(k)  \Big\}  - \Tr\Big\{ \wtg_0 S(p)\wtg^3 k_z \Big(k_0 + \frac{\Omega\Sigma_3}{2}\Big)S(k) \Big\} \nn \\ && + \Tr\Big\{ \wtg_0 p_z S(p)\wtg_3\Big(k_0 + \frac{\Omega\Sigma_3}{2}\Big)S(k) \Big\}  + \Tr\Big\{ \wtg^3 p_z S(p)\wtg^0 \Big(k_0 + \frac{\Omega\Sigma_3}{2}\Big) S(k) \Big\}\nn \\ && - \Tr\Big\{ \wtg^3 p_z \Big(p_0 + \frac{\Omega\Sigma_3}{2}\Big)S(p)\wtg_0 S(k)  \Big\}  -  \Tr\Big\{ \wtg_3 S(p)\wtg_0 k_z\Big( k_0 + \frac{\Omega\Sigma_3}{2}\Big)S(k) \Big\} \nn \\ && - \Tr\Big\{ \wtg_3\Big(p_0 + \frac{\Omega\Sigma_3}{2}\Big)S(p)\wtg_0 k_z S(k) \Big\}   -(p_z + k_z)^2\Tr\Big\{ \wtg^3S(p)\wtg^3S(k)  \Big\}  \nn \\ &&  + \Tr\Big\{ \wtg^0\Big(p_0 + \frac{\Omega\Sigma_3}{2}\Big)S(p) \wtg^0\Big(k_0 + \frac{\Omega\Sigma_3}{2}\Big)S(k) \Big\}   -\Tr\Big\{ \wtg^0\Big(p_0 + \frac{\Omega\Sigma_3}{2}\Big)^2S(p)\wtg^0S(k)  \Big\} \nn \\ && -\Tr\Big\{ \wtg^0S(p)\wtg^0\Big(k_0 + \frac{\Omega\Sigma_3}{2}\Big)^2S(k) \Big\}   - \Tr\Big\{\wtg^3p_zS(p)\wtg^0\Big(k_0 + \frac{\Omega\Sigma_3}{2}\Big)S(k)\Big\} \nn \\ && + \Tr\Big\{ \wtg^3p_z\Big(p_0 + \frac{\Omega\Sigma_3}{2}\Big)S(p)\wtg_0S(k)  \Big\} - \Tr\Big\{ \wtg_3S(p)\wtg^0k_z\Big(k_0 + \frac{\Omega\Sigma_3}{2}\Big)S(k)  \Big\} \nn \\ && + \Tr\Big\{ \wtg_3 \Big( p_0 + \frac{\Omega\Sigma_3}{2} \Big)S(p)\wtg_0k_zS(k) \Big\}   - \Tr\Big\{  \wtg_0 \Big( p_0 + \frac{\Omega\Sigma_3}{2}  \Big)^2S(p)\wtg_0 S(k) \Big\}\nn \\ && - \Tr\Big\{ \wtg_0S(p)\wtg^0\Big(k_0 + \frac{\Omega\Sigma_3}{2}\Big)^2S(k)  \Big\}   + \Tr\Big\{ \wtg_0 \Big(p_0 + \frac{\Omega\Sigma_3}{2}\Big)S(p)\wtg_0 \Big(k_0 + \frac{\Omega\Sigma_3}{2}\Big)S(k)  \Big\} \nn \\ && + \Tr\Big\{ \wtg_0\Big(p_0 + \frac{\Omega\Sigma_3}{2}\Big)S(p)\wtg^3k_zS(k)  \Big\}  + \Tr\Big\{ \wtg_0\Big(p_0 + \frac{\Omega\Sigma_3}{2}\Big)p_zS(p)\wtg_3S(k)  \Big\}\nn \\ && - \Tr\Big\{\wtg_0S(p)\wtg^3k_z\Big(k_0  + \frac{\Omega\Sigma_3}{2}\Big)S(k) \Big\} -\Tr\Big\{ \wtg_0 p_zS(p)\wtg_3\Big(k_0 + \frac{\Omega\Sigma_3}{2}\Big)S(k)  \Big\} \nn \\ && - \Tr\Big\{\wtg_0\Big(p_0 + \frac{\Omega\Sigma_3}{2}\Big)S(p)\wtg^0\Big(k_0 + \frac{\Omega\Sigma_3}{2}\Big)S(k)  \Big\} + \Tr\Big\{ \wtg_0\Big(p_0 + \frac{\Omega\Sigma_3}{2}\Big)^2S(p)\wtg^0S(k)\Big\}\nn \\ &&  - \Tr\Big\{ \wtg_0S(p)\wtg^0\Big(k_0 + \frac{\Omega\Sigma_3}{2}\Big)^2S(k) \Big\}  \Bigg\}\Bigg]\,.	
 \label{spec-10}
	\eea
 \end{widetext}
Upon integrating over $d^4 r$ and $d^4k$, we obtain a delta function $(2\pi)^4\delta(q - p - k)$, ensuring momentum conservation at the vertex. This conservation sets specific relationships between $p$, $k$, and $q$: $k_0 = q_0 - p_0$, $k_z = q_z - p_z$, and $k_\perp = q_\perp - p_\perp$, consistent with the one-loop diagram for photon polarization shown in Fig.~\ref{fig:oneloop}. The spectral function, Eq.~(\ref{spec-1A}), is then computed using Eq.~(\ref{spec-10}) and is expressed as
\begin{widetext}
    \bea
	\rho_\eta(q)  &=& \text{Im~}i\int d^4r~ e^{iq\cdot r}\big\langle  \Delta^{\mu\nu\alpha\beta}T_{\alpha\beta}   \Delta_{\mu\nu\rho\lambda}T^{\rho\lambda}  \big\rangle_R \nn \\
	&=& \text{Im}\frac{i}{40}\int d^4r\int\frac{d^4p}{(2\pi)^4} \Bigg[  (q_z - 2p_z)^2\Big\{ \Tr\big\{\wtg_1 S(p)\wtg^1 S(q-p) + \wtg_3 S(p)\wtg^3 S(q-p)\big\}    \Big\} \nn \\ &&+ \frac{1}{3} \Bigg\{   -2 \Tr\Big\{\wtg^0\Big(p_0 + \frac{\Omega\Sigma_3}{2}\Big)S(p)\wtg^0 \Big(q_0 - p_0 + \frac{\Omega\Sigma_3}{2}\Big) S(q-p)\Big\}  + \Tr\Big\{ \wtg^0\Big(p_0 + \frac{\Omega\Sigma_3}{2}\Big)^2 S(p) \wtg_0 S(q-p)\Big\} \nn \\ &&  +  \Tr\Big\{ \wtg^0 S(p) \wtg_0 \Big(q_0 - p_0 + \frac{\Omega\Sigma_3}{2}\Big)^2S(q-p)\Big\}  + \Tr\Big\{ \wtg^0\Big(p_0 + \frac{\Omega\Sigma_3}{2}\Big)S(p)\wtg^3 k_z S(q-p)  \Big\} \nn \\ && + \Tr\Big\{ \wtg^0p_z\Big(p_0 + \frac{\Omega\Sigma_3}{2}\Big)S(p)\wtg_3S(q-p)  \Big\}  - \Tr\Big\{ \wtg_0 S(p)\wtg^3 (q_z - p_z) \Big(q_0 - p_0 + \frac{\Omega\Sigma_3}{2}\Big)S(q-p) \Big\} \nn \\ && + \Tr\Big\{ \wtg_0 p_z S(p)\wtg_3\Big(q_0 - p_0 + \frac{\Omega\Sigma_3}{2}\Big)S(q-p) \Big\}  + \Tr\Big\{ \wtg^3 p_z S(p)\wtg^0 \Big(q_0 - p_0 + \frac{\Omega\Sigma_3}{2}\Big) S(q-p) \Big\}\nn \\ && - \Tr\Big\{ \wtg^3 p_z \Big(p_0 + \frac{\Omega\Sigma_3}{2}\Big)S(p)\wtg_0 S(q-p)  \Big\}  -  \Tr\Big\{ \wtg_3 S(p)\wtg_0( q_z - p_z)\Big( q_0 - p_0 + \frac{\Omega\Sigma_3}{2}\Big)S(q-p) \Big\} \nn \\ && - \Tr\Big\{ \wtg_3\Big(p_0 + \frac{\Omega\Sigma_3}{2}\Big)S(p)\wtg_0( q_z - p_z)S(q-p) \Big\}   -q_z^2\Tr\Big\{ \wtg^3S(p)\wtg^3S(q-z)  \Big\}  \nn \\ &&  + \Tr\Big\{ \wtg^0\Big(p_0 + \frac{\Omega\Sigma_3}{2}\Big)S(p) \wtg^0\Big(q_0 - p_0 + \frac{\Omega\Sigma_3}{2}\Big)S(q-p) \Big\}   -\Tr\Big\{ \wtg^0\Big(p_0 + \frac{\Omega\Sigma_3}{2}\Big)^2S(p)\wtg^0S(q-p)  \Big\} \nn \\ && -\Tr\Big\{ \wtg^0S(p)\wtg^0\Big(q_0 - p_0 + \frac{\Omega\Sigma_3}{2}\Big)^2S(q-p) \Big\}   - \Tr\Big\{\wtg^3p_zS(p)\wtg^0\Big(q_0 - p_0 + \frac{\Omega\Sigma_3}{2}\Big)S(q-p)\Big\} \nn \\ && + \Tr\Big\{ \wtg^3p_z\Big(p_0 + \frac{\Omega\Sigma_3}{2}\Big)S(p)\wtg_0S(q-p)  \Big\} - \Tr\Big\{ \wtg_3S(p)\wtg^0(q_z - p_z)\Big(q_0 - p_0 + \frac{\Omega\Sigma_3}{2}\Big)S(q-p)  \Big\} \nn \\ && + \Tr\Big\{ \wtg_3 \Big( p_0 + \frac{\Omega\Sigma_3}{2} \Big)S(p)\wtg_0(q_z - p_z)S(q-p) \Big\}   - \Tr\Big\{  \wtg_0 \Big( p_0 + \frac{\Omega\Sigma_3}{2}  \Big)^2S(p)\wtg_0 S(q-p) \Big\}\nn \\ && - \Tr\Big\{ \wtg_0S(p)\wtg^0\Big(q_0 - p_0 + \frac{\Omega\Sigma_3}{2}\Big)^2S(q-p)  \Big\}   + \Tr\Big\{ \wtg_0 \Big(p_0 + \frac{\Omega\Sigma_3}{2}\Big)S(p)\wtg_0 \Big(q_0 - p_0 + \frac{\Omega\Sigma_3}{2}\Big)S(q-p)  \Big\} \nn \\ && + \Tr\Big\{ \wtg_0\Big(p_0 + \frac{\Omega\Sigma_3}{2}\Big)S(p)\wtg^3(q_z - p_z)S(q-p)  \Big\}  + \Tr\Big\{ \wtg_0\Big(p_0 + \frac{\Omega\Sigma_3}{2}\Big)p_zS(p)\wtg_3S(q-p)  \Big\}\nn \\ && - \Tr\Big\{\wtg_0S(p)\wtg^3(q_z - p_z)\Big(q_0 - p_0  + \frac{\Omega\Sigma_3}{2}\Big)S(q-p) \Big\} -\Tr\Big\{ \wtg_0 p_zS(p)\wtg_3\Big(q_0 - p_0 + \frac{\Omega\Sigma_3}{2}\Big)S(q-p)  \Big\} \nn \\ && - \Tr\Big\{\wtg_0\Big(p_0 + \frac{\Omega\Sigma_3}{2}\Big)S(p)\wtg^0\Big(q_0 - p_0 + \frac{\Omega\Sigma_3}{2}\Big)S(q-p)  \Big\} + \Tr\Big\{ \wtg_0\Big(p_0 + \frac{\Omega\Sigma_3}{2}\Big)^2S(p)\wtg^0S(q-p)\Big\}\nn \\ &&  - \Tr\Big\{ \wtg_0S(p)\wtg^0\Big(q_0 - p_0 + \frac{\Omega\Sigma_3}{2}\Big)^2S(q-p) \Big\}  \Bigg\}\Bigg]\,.	
 \label{spec-11}
	\eea
\end{widetext}
To execute the integration in Eq.~(\ref{spec-11}), we will carry out the Matsubara frequency summation, considering finite angular velocity and chemical potential. In the following we present details the Matsubara frequency summation process applied to compute the spectral function of energy-momentum tensors. At finite temperature
\bea	
 p_0 &\to& i\widetilde{\omega}_N = (2N + 1)\pi T\,, \quad q_0 \to i\nu_N = 2\pi NT\,,\nn\\
 \int\frac{d^4p}{(2\pi)^4}&\equiv& \sumintof = iT\sum_{N = -\infty}^{+\infty}\int\frac{d^3p}{(2\pi)^3},
 \eea
where $N\in \mathbb{Z}$,
and 
$p_0$ and $q_0$ are the temporal components of the four-momentum of the fermion and boson, respectively, in the photon polarization. In one-loop calculations of spectral functions, the fermion propagator under rotation, Eq.~\eqref{F-Rot-8}, leads to fermionic Matsubara frequency summations of the type
\begin{widetext}
	\bea
	(a) &\qquad& T\sum_{\{p_N\}}\frac{1}{\big[ i\widetilde{\omega}_N + \frac{\Omega}{2} + \mu  \big]^2 - \vec{p}^2 - m^2}    \frac{1}{\big[ i\nu_N - i\widetilde{\omega}_N + \frac{\Omega}{2} - \mu  \big]^2 - \vec{k}^2 - m^2}\,,
	\label{spec-12-A}\\
	(b) &\qquad& T\sum_{\{p_N\}}\frac{1}{\big[ i\widetilde{\omega}_N + \frac{\Omega}{2} + \mu  \big]^2 - \vec{p}^2 - m^2}    \frac{1}{\big[ i\nu_N - i\widetilde{\omega}_N - \frac{\Omega}{2} - \mu  \big]^2 - \vec{k}^2 - m^2}\,,
	\label{spec-12-B}\\
	(c) &\qquad& T\sum_{\{p_N\}}\frac{1}{\big[ i\widetilde{\omega}_N - \frac{\Omega}{2} + \mu  \big]^2 - \vec{p}^2 - m^2}    \frac{1}{\big[ i\nu_N - i\widetilde{\omega}_N + \frac{\Omega}{2} - \mu  \big]^2 - \vec{k}^2 - m^2}\,,
	\label{spec-12-C}\\
	(d) &\qquad& T\sum_{\{p_N\}}\frac{1}{\big[ i\widetilde{\omega}_N - \frac{\Omega}{2} + \mu  \big]^2 - \vec{p}^2 - m^2}    \frac{1}{\big[ i\nu_N - i\widetilde{\omega}_N - \frac{\Omega}{2} - \mu  \big]^2 - \vec{k}^2 - m^2}\,.
	\label{spec-12-D}
	\eea 
\end{widetext}	
Utilizing the Saclay method for evaluating Matsubara frequency sums~\cite{Laine:2016hma}, we derive the results for the frequency summations, Eqs.~(\ref{spec-12-A})-(\ref{spec-12-D}), expressed as
	\begin{widetext}
		\bea
		&&(a) \quad T\sum_{\{p_N\}}\frac{1}{\big[ i\widetilde{\omega}_N + \frac{\Omega}{2} + \mu  \big]^2 - \vec{p}^2 - m^2}    \frac{1}{\big[ i\nu_N - i\widetilde{\omega}_N + \frac{\Omega}{2} - \mu  \big]^2 - \vec{k}^2 - m^2} \nn \\
		&&= \frac{1}{4E_pE_k}\Bigg\{  \frac{n_F(E_p + \mu + \Omega/2) + n_F(E_k - \mu + \Omega/2) - 1}{q_0 - E_p - E_k} + \frac{n_F(E_k + \mu + \Omega/2) - n_F(E_p + \mu + \Omega/2)}{q_0 + E_k - E_p}  \nn \\
		&&+ \frac{n_F(E_p - \mu - \Omega/2) - n_F(E_k - \mu - \Omega/2)}{q_0 + E_p - E_k} + \frac{1 - n_F(E_p - \mu - \Omega/2) - n_F(E_k + \mu - \Omega/2)}{q_0 + E_k + E_p} \Bigg\}\,,
		\label{spec-12-E}
		\eea 
		\bea
		&&(b) \quad T\sum_{\{p_N\}}\frac{1}{\big[ i\widetilde{\omega}_N + \frac{\Omega}{2} + \mu  \big]^2 - \vec{p}^2 - m^2}    \frac{1}{\big[ i\nu_N - i\widetilde{\omega}_N - \frac{\Omega}{2} - \mu  \big]^2 - \vec{k}^2 - m^2} \nn \\
		&&= \frac{1}{4E_pE_k}\Bigg\{  \frac{n_F(E_p + \mu + \Omega/2) + n_F(E_k - \mu - \Omega/2) - 1}{q_0 - E_p - E_k} + \frac{n_F(E_k + \mu + \Omega/2) - n_F(E_p + \mu + \Omega/2)}{q_0 + E_k - E_p}  \nn \\
		&&+ \frac{n_F(E_p - \mu - \Omega/2) - n_F(E_k - \mu - \Omega/2)}{q_0 + E_p - E_k} + \frac{1 - n_F(E_p - \mu - \Omega/2) - n_F(E_k + \mu + \Omega/2)}{q_0 + E_k + E_p} \Bigg\}\,,
		\label{spec-12-F}
		\eea 
		\bea
		&&(c) \quad T\sum_{\{p_N\}}\frac{1}{\big[ i\widetilde{\omega}_N - \frac{\Omega}{2} + \mu  \big]^2 - \vec{p}^2 - m^2}    \frac{1}{\big[ i\nu_N - i\widetilde{\omega}_N + \frac{\Omega}{2} - \mu  \big]^2 - \vec{k}^2 - m^2} \nn \\
		&&= \frac{1}{4E_pE_k}\Bigg\{  \frac{n_F(E_p + \mu - \Omega/2) + n_F(E_k - \mu + \Omega/2) - 1}{q_0 - E_p - E_k} + \frac{n_F(E_k + \mu - \Omega/2) - n_F(E_k + \mu - \Omega/2)}{q_0 + E_k - E_p}  \nn \\
		&&+ \frac{n_F(E_p - \mu + \Omega/2) - n_F(E_k - \mu + \Omega/2)}{q_0 + E_p - E_k} + \frac{1 - n_F(E_p - \mu + \Omega/2) - n_F(E_k + \mu - \Omega/2)}{q_0 + E_k + E_p} \Bigg\}\,,
		\label{spec-12-G}
		\eea
		\bea
		&&(d) \quad T\sum_{\{p_N\}}\frac{1}{\big[ i\widetilde{\omega}_N - \frac{\Omega}{2} + \mu  \big]^2 - \vec{p}^2 - m^2}    \frac{1}{\big[ i\nu_N - i\widetilde{\omega}_N - \frac{\Omega}{2} - \mu  \big]^2 - \vec{k}^2 - m^2} \nn \\
		&&= \frac{1}{4E_pE_k}\Bigg\{  \frac{n_F(E_p + \mu - \Omega/2) + n_F(E_k - \mu - \Omega/2) - 1}{q_0 - E_p - E_k} + \frac{n_F(E_k + \mu - \Omega/2) - n_F(E_k + \mu - \Omega/2)}{q_0 + E_k - E_p}  \nn \\
		&&+ \frac{n_F(E_p - \mu + \Omega/2) - n_F(E_k - \mu + \Omega/2)}{q_0 + E_p - E_k} + \frac{1 - n_F(E_p - \mu + \Omega/2) - n_F(E_k + \mu - \Omega/2)}{q_0 + E_k + E_p} \Bigg\}.
		\label{spec-12-H}
		\eea
	\end{widetext}
In the above equations, there are four different combinations of $q_0$, $E_k$, and $E_p$ in the denominators: $q_0 \pm E_k \pm E_p$ and $q_0 \mp E_k \pm E_p$, which can be interpreted as energy limit values from the upper half plane. By transforming $\nu_N \to -i(q_0 + i0^+)$ through the analytic continuation of discrete Matsubara frequencies to continuous energies, we can extract the imaginary part as needed in Eq.~(\ref{spec-11}), utilizing the relation
	\bea
	\frac{1}{\Delta \pm i0^+} = \mathbb{P}\left(\frac{1}{\Delta}\right) \mp i\pi \delta(\Delta).
	\label{spec-13}
	\eea  
where $\mathbb{P}$ is the principal part of the function. Employing Eq.~(\ref{spec-13}), $\rho_\eta(q)$ is given as
 \begin{widetext}
	\bea
	\rho_\eta(q) &=& -\pi \int\frac{d^3p}{(2\pi)^3}\Big(\frac{ \mathcal{C}(p_\perp, p_z, p_0, q_\perp, q_z, q_0)}{4E_pE_{q-p}}\Big) \Bigg[\Big\{n_F(E_{q-p} + \bar{\mu}) - n_F(E_p + \bar{\mu})\Big\}\Big\{\delta(q_0 + E_{q-p} - E_p)\Big\}\nn \\
	&& +\, \Big\{n_F(E_p - \bar{\mu}) - n_F(E_{q-p} - \bar{\mu})\Big\}\Big\{\delta(q_0 + E_p - E_{q-p})\Big\} \nn \\ && +\,  \Big\{ n_F(E_{q-p} + \bar{\mu} ) + n_F(E_p - \bar{\mu}) -1  \Big\}\Big\{\delta(q_0 - E_{q-p} - E_p)\Big\} \nn \\
	&& +\, \Big\{ n_F(E_{q-p} - \bar{\mu} ) + n_F(E_p + \bar{\mu}) -1  \Big\}\Big\{\delta(q_0 + E_{q-p} + E_p)\Big\}  \Bigg] \nn \\
	&&-\, \pi \int\frac{d^3p}{(2\pi)^3}\Big(\frac{\mathcal{D}(p_\perp, p_z, p_0, q_\perp, q_z, q_0)}{4E_pE_{q-p}}\Big) \Bigg[\Big\{n_F(E_{q-p} + \widetilde{\mu}) - n_F(E_p + \widetilde{\mu})\Big\}\Big\{\delta(q_0 + E_{q-p} - E_p)\Big\}\nn \\
	&& +\, \Big\{n_F(E_p - \widetilde{\mu}) - n_F(E_{q-p} - \widetilde{\mu})\Big\}\Big\{\delta(q_0 + E_p - E_{q-p})\Big\} \nn \\ && +\,  \Big\{ n_F(E_{q-p} + \widetilde{\mu} ) + n_F(E_p - \widetilde{\mu}) -1  \Big\}\Big\{\delta(q_0 - E_{q-p} - E_p)\Big\} \nn \\
	&& +\, \Big\{ n_F(E_{q-p} - \widetilde{\mu} ) + n_F(E_p + \widetilde{\mu}) -1  \Big\}\Big\{\delta(q_0 + E_{q-p} + E_p)\Big\}  \Bigg]\,,
	\label{spec-14}
	\eea 
 \end{widetext}
where $\bar{\mu} = \mu + \frac{\Omega}{2}\text{~~and~~}\widetilde{\mu} = \mu - \frac{\Omega}{2}$ are the combinations of chemical potential and angular velocity. $\mathcal{C}(p_\perp, p_z, p_0, q_\perp, q_z, q_0)$ and $\mathcal{D}(p_\perp, p_z, p_0, q_\perp, q_z, q_0) $ are functions expressed as
\begin{widetext}
\bea 
&&\mathcal{C}(p_\perp, p_z, p_0, q_\perp, q_z, q_0) = -\frac{1}{480}\Big\{ 2(p_z - q_z)\big(q_0 - p_0 + 8p_z - 2p_0 + \Omega\big) + 8(q_z - p_z)^2 + 3(q_0 - p_0) \Omega \nn \\&& +\, p_z\big(8p_z - 2p_0 + \Omega\big)  \big]  
\big[(q_z - p_z)\big( 4p_z + 4p_0 - 2\Omega \big)- 4(q_\perp - p_\perp )p_\perp + \big(2q_0 - 2p_0- \Omega\big)\big(2p_z + 2p_0 - \Omega\big) \big]\nn \\ &&+ \frac{1}{10}\big[  p_z\big( q_z - 2p_z \big)   \big]\big[(q_z-p_z)\big(4p_z - 4p_0 - 2\Omega \big) - 4(q_\perp - p_\perp) p_\perp + \big(2q_0 - 2p_0 - \Omega\big)\big( 2p_z + 2p_0 + \Omega \big)  \Big\}\,,
	\label{spec-14-A} 
 \eea
 \bea
 &&\mathcal{D}(p_\perp, p_z, p_0, q_\perp, q_z, q_0) =  -\frac{1}{480} \Big\{ 2(p_z - q_z)\big(q_0 - p_0 + 8p_z + 2p_0 + \Omega\big) + 8(q_z - p_z)^2 + 3(q_0 - p_0)\Omega \nn \\ && +\, p_z\big(8p_z + 2p_0 + \Omega\big)  \big]  \big[-k_z\big( -4p_z + 4p_0 + 2\Omega \big)- 4(q_\perp - p_\perp) p_\perp + \big(2q_0 - 2p_0 + \Omega\big)\big(-2p_z + 2p_0 + \Omega\big) \big] \nn \\ && + \frac{1}{10}\big[  p_z\big( q_z - 2p_z \big)   \big]\big[(q_z- p_z)\big(4p_z + 4p_0 + 2\Omega \big) - 4(q_\perp - p_\perp)p_\perp + \big(2q_0 - 2p_0 + \Omega\big)\big( -2p_z + 2p_0 + \Omega \big)  \Big\}.
 \label{spec-14-B}
	\eea 
 \end{widetext}
It's important to note that $\rho_\eta(q)$, as determined in Eq.~(\ref{spec-14}) through analytic continuation and the cut of a Euclidean correlator, represents real scatterings of on-shell particles. These scatterings are kinematically allowed and their distribution is described by the Fermi-Dirac distribution function. More precisely, the imaginary parts are associated with real particle scatterings, information about which is conveyed by the delta functions in Eq.~(\ref{spec-14}). The delta functions $\delta(q_0 \pm E_p \mp E_k)$, known as Landau cuts, and $\delta(q_0 \pm E_p \pm E_k)$, known as unitary cuts. Landau cuts only occur at finite temperature, while unitary cuts, also present due to vacuum contributions, don't contribute to transport coefficient calculations and are thus omitted here~\cite{Mallik:2016anp}. From Eq.~(\ref{spec-1}), the shear viscosity, as calculated using Eq.~(\ref{spec-14}), is
\begin{widetext}
	\bea
	\eta &=& \displaystyle{\lim_{ q_0 \to 0, \,\vec{q}=\vec{0}}}\text{Im}\Bigg[-\pi\int\frac{d^3p}{(2\pi)^3}\frac{\mathcal{C}(p_\perp, p_z, p_0, q_\perp , q_z , q_0 )}{4E_{q-p}E_p}\nn \\ &&\displaystyle{\lim_{\Gamma\to 0}}\Bigg\{\frac{\big(n_F(E_{q-p} + \bar{\mu}) - n_F(E_p + \bar{\mu})\big)/q_0}{q_0 + E_{q-p} - E_p + i\Gamma}  + \frac{\big(n_F(E_{q-p} - \bar{\mu}) - n_F(E_p - \bar{\mu})\big)/q_0}{q_0 + E_p - E_{q-p} + i\Gamma} \Bigg\}\Bigg]\nn \\
&+&  \displaystyle{\lim_{q_0 \to 0, \,\vec{q}=\vec{0}}}\text{Im}\Bigg[-\pi\int\frac{d^3p}{(2\pi)^3}\frac{\mathcal{D}(p_\perp, p_z, p_0, q_\perp, q_z , q_0 )}{4E_{q-p}E_p}\nn \\ &&\displaystyle{\lim_{\Gamma\to 0}}\Bigg\{\frac{\big(n_F(E_{q-p} + \widetilde{\mu}) - n_F(E_p + \widetilde{\mu})\big)/q_0}{q_0 + E_{q-p} - E_p + i\Gamma} + \frac{\big(n_F(E_{q-p} - \widetilde{\mu}) - n_F(E_p - \widetilde{\mu})\big)/q_0}{q_0 + E_p - E_{q-p} + i\Gamma} \Bigg\}\Bigg]\,.
	\label{spec-15}
	\eea 
 \end{widetext}
In the equation, we use the representation $\delta(x) = -(1/\pi)\lim_{\Gamma\to 0}\frac{1}{x + i\Gamma}$, where $\Gamma$ is the thermal width of the medium. This width arises from interactions within the medium at finite temperature, with high system temperatures effectively generating interactions (i.e., a finite $\Gamma$). Notably, $\Gamma$ is essentially the inverse of the relaxation time $\tau$, i.e., $\Gamma = 1/\tau$~\cite{Mallik:2016anp}. 
In this work, $\Gamma$ is introduced via the Breit-Wigner form, and a similar approach can be applied by including a half-width $i\Gamma/2$~\cite{Lang:2012tt} in the momentum space propagators. This involves considering the expression for the inverse of the retarded momentum space propagator $G^{-1}_{\text{R}}(p)$ with the complex self-energy $\Sigma_\text{R}(p)$: 
$G_{\text{R}}^{-1}(p) = p^2 - m_0^2 - \text{Re}\Sigma_{\text{R}}(p) - i\text{Im}\Sigma_{\text{R}}(p)\sim (p_0 + i\gamma(p))^2 - E_p^2$, where $\gamma = -\frac{1}{2p_0}\text{Im}\Sigma_{\text{R}}(p) = \frac{\Gamma}{2}$. This indicates that $\Gamma$ is inherent in the theory's propagator. In our calculations, we use a specific form of the $\delta(x)$ function to incorporate this information. $\Gamma$ can be calculated using an interacting Lagrangian and thermal field theory techniques~\cite{Laine:2016hma,Lang:2012tt}. At the $\vec{q} =\vec{0},\, q_0 \to 0$ limit, $\eta$ appears indeterminate (0/0 form), which is unphysical. To address this, we apply L'Hospital's rule. In a free theory where interaction is absent ($\Gamma \to 0$), shear viscosity becomes infinite. Thus, a finite $\Gamma$ is necessary to yield a non-divergent shear viscosity. The final expression for $\eta(T,\mu,\Omega)$ is then
 \begin{widetext}	
\bea
&&\eta (T,\mu,\Omega) = \frac{1}{240}\int\frac{d^3p}{(2\pi)^3}\left(\frac{1}{4E_p^2\Gamma T}\right)
\label{spec-15A}\\
&&\Bigg\{\Big\{160p_z^4 + 4p_\perp^2\Omega^2 + \big(\Omega - E_p- 3p_z\big)\Omega^3 + 8p_\perp^2p_z\big( \Omega -4E_p \big)  + 32p_z^3\big( \Omega + 5E_p \big) - 8p_z^2\Omega\big( 2E_p + 3\Omega  \big) \Big\} \mathcal{N}_{\pm\mu, \mp\Omega/2} \nn \\
&&+  \Big\{160p_z^4  + 4p_\perp^2\Omega^2 + \big(\Omega - E_p- 3p_z\big)\Omega^3 +  8p_\perp^2p_z\big( \Omega -4E_p \big)+ 32p_z^3\big(\Omega - 5 E_p\big) - 8p_z^2\Omega\big( 2E_p- 3\Omega \big) \Big\} \mathcal{N}_{\pm\mu, \pm\Omega/2}\Bigg\}\,,\nn
	\eea 
where
\bea
&&\mathcal{N}_{\pm\mu, \pm\Omega/2} = \sum_{s = \pm 1}n_F(E_p + s\mu + s\Omega/2)\big\{1 - n_F(E_p + s\mu + s\Omega/2)\big\} \nn \\
&&\mathcal{N}_{\pm\mu, \mp\Omega/2} = \sum_{\lambda = \pm 1}n_F(E_p + \lambda\mu - \lambda\Omega/2)\big\{1 - n_F(E_p + \lambda\mu - \lambda\Omega/2)\big\},
\label{spec-16}
\eea 
 \end{widetext}
and $n_F(x)$ is the Fermi-Dirac distribution function. Equation~(\ref{spec-15A}) provides the formula for calculating the shear viscosity in a rotating, hot, and dense fermionic system. 
This formula is particularly relevant for a hadronic medium consisting of spin-$1/2$ particles. However, it's crucial to maintain the magnitude of $\Omega$ at or above $0.1$ GeV to apply this result effectively. In Eq.~(\ref{spec-15A}), the influence of $\Omega$ is incorporated both in the Fermi-Dirac distribution function and the numerator of the expression. This indicates that $\Omega$ functions similarly to an effective chemical potential within the medium.
\section{Results}
\label{Res}
In this section, we present the numerical results on the temperature ($T$), chemical potential ($\mu$), and angular velocity ($\Omega$) dependence of the shear viscosity ($\eta$) for a system of spin-$1/2$ fermions subjected to a high angular frequency. The quantum field theoretical approach to studying $\eta$ allows us to explore the roles of $\mu$ and its effective counterpart $\Omega$ from a first-principles perspective. Both $\mu$ and $\Omega$ enter the distribution function via Matsubara frequency summation, but $\Omega$ also influences the momentum space propagator, as evident from Eq.~(\ref{F-Rot-8}), acting as an effective, rather than exact, chemical potential.

We consider temperatures in the range of $0.15 - 0.6$ GeV, approximately the regime where deconfined QCD matter exists, with $0.6$ GeV being the upper limit of achievable temperature. In this deconfined state, we explore $\mu$ values of $0.05$, $0.2$, and $0.4$ GeV. Given the focus on large angular velocity, $\Omega$ is set in the order of $0.1-1$ GeV.

It's crucial to note that $\Omega$ cannot be excessively large for two key reasons. Firstly, a very high $\Omega$ would breach causality, disrupting global equilibrium. For thermal equilibrium under rotation, the local temperature $T$ should vary so that $T/\gamma$ ($\gamma$ being the time dilation factor, where $\gamma = 1/\sqrt{(1 - \Omega^2r^2)}$ and $r$ the cylinder radius) remains constant. Hence, $\Omega < 1/r$ is necessary for valid Lorentz transformation, preventing $T$ from becoming imaginary. For $\Omega = 1/r$, $T$ approaches infinity, an unphysical scenario. Therefore, $\Omega$ must be balanced with other system energy scales like $T$ and $\mu$ to maintain equilibrium.

Secondly, as reported in Ref.~\cite{Braguta:2023yjn}, rotational instabilities can occur in a rotating QGP. Here, the moment of inertia is negative below a super-vortical temperature ($T_s = 1.5 T_c$, with $T_c$ being the deconfining transition temperature for non-rotating plasma), akin to findings for spinning Kerr and Myers-Perry black holes. This negative moment of inertia depends on the system's velocity ($v = \Omega r$), with $r$ indicating system size. Since $\Omega$, $T$, $\mu$, and $\Lambda_{\rm QCD}$ are interdependent, care is needed in handling thermodynamic and transport quantities to avoid negative values from rotational instabilities.

Our work aims to demonstrate the interplay between $\eta$ and $\Omega$, with $\eta(T,\mu, \Omega$) showing a complex dependence on $\Omega$ (see Eq.~(\ref{spec-15A})). Therefore, a thoughtful approach is required to manage $\Omega$'s range without impacting the system's physical properties.
\subsection*{Chemical potential \texorpdfstring{$\mu$}{} vs Effective chemical potential~\texorpdfstring{$\Omega$}{}}
\begin{figure}[ht!]
\includegraphics[scale = 0.24]{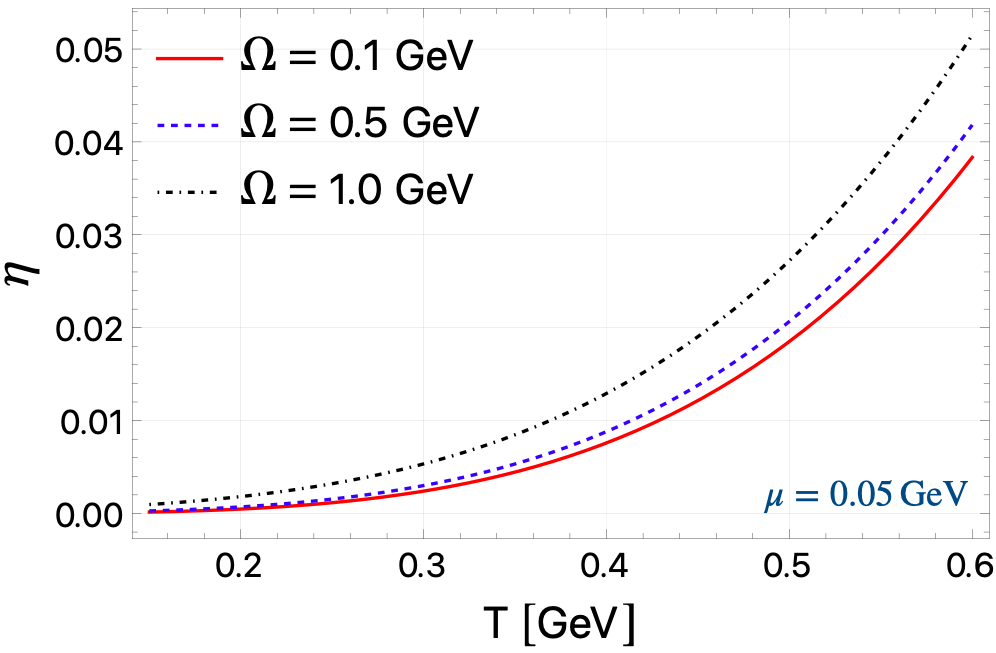}
\caption{Behaviour of $\eta$ with $T$ for $\Omega$ = 0.1, 0.5, 1.0 GeV at $\mu$ = 0.05 GeV.}
\label{fig-T1}
\end{figure}
\begin{figure}[ht!]
\includegraphics[scale = 0.24]{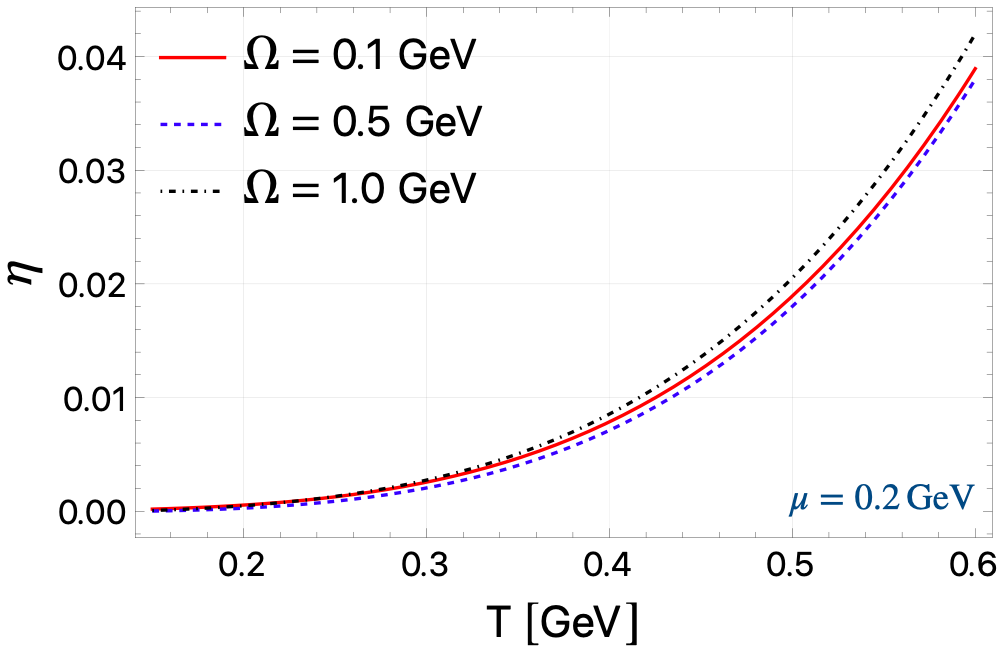}
\caption{Same as fig.~\ref{fig-T1} but for $\mu$ = 0.2 GeV.}
\label{fig-T2}
\end{figure}
\begin{figure}[ht!]
\includegraphics[scale = 0.24]{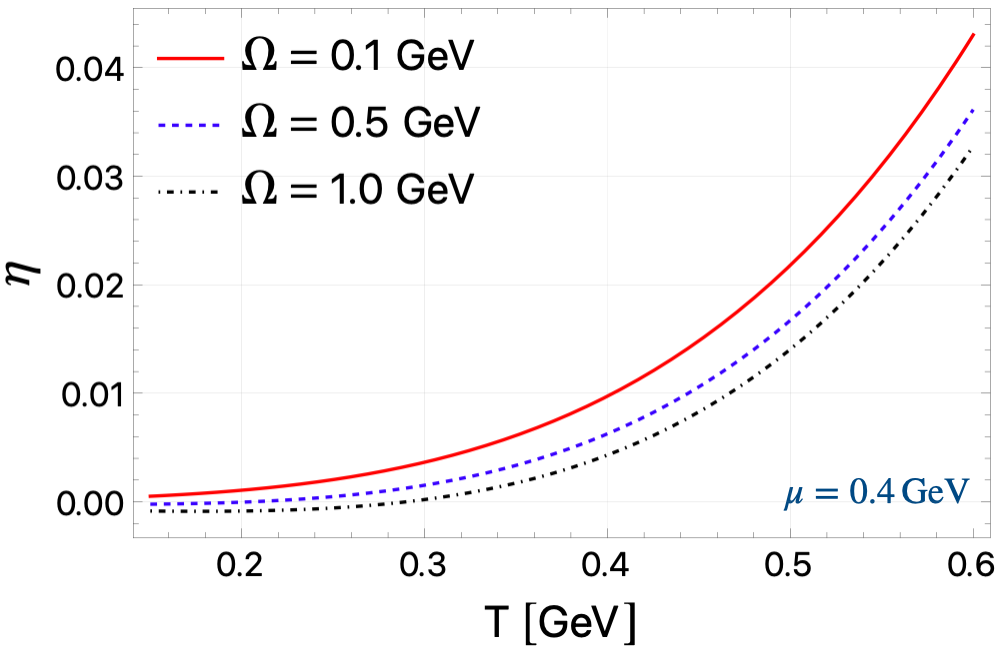}
\caption{Same as fig.~\ref{fig-T1} but for  $\mu$ = 0.4 GeV.}
\label{fig-T3}
\end{figure}
Our investigation predominantly characterizes QGP through its shear viscosity, although both shear and bulk viscosities are crucial. For a more comprehensive understanding of QGP's phenomenological properties, it's essential to focus on specific viscosities like shear viscosity to entropy density ratio ($\eta/s$) and bulk viscosity to entropy density ratio ($\zeta/s$). Our study provides a universal framework for examining the shear viscosity ($\eta$) in fermionic systems. It's adaptable for exploring the dynamics of quarks and spin-$1/2$ baryons, offering insights into the broader scope of QGP properties.

In our study, as depicted in Figures~\ref{fig-T1} to \ref{fig-T3}, we analyze the variation of $\eta$ with $T$ across different $\mu$. In Figure~\ref{fig-T1}, focusing on $\mu = 0.05$ GeV, $\eta$ is examined over a temperature range from 0.15 to 0.6 GeV for various angular velocities ($\Omega$ values of 0.1, 0.5, and 1 GeV). This analysis reveals that $\eta$ typically increases with temperature, a characteristic behavior of relativistic fluids. Notably, $\eta$ also shows an increase with $\Omega$, particularly in the lower temperature range. This trend aligns with expectations, but may not hold for smaller $\Omega$ values (below 0.1 GeV), considering the limitations of our employed propagator, which is more accurate for larger $\Omega$ values.
The increasing nature of $\eta$ with $\Omega$ may be deduced mathematically from Eq.~(\ref{spec-15A}), where the expression shows that there is a $\mathcal{O}(\Omega^4)$ dependence of $\eta$ on $\Omega$. Additionally, in a rotating fluid, the shear flow induced by velocity gradients is further amplified by the rotation, as the velocity $v = \Omega r$ and its gradient $\vec{\nabla} v \sim \mathcal{O}(\Omega)$ introduce an extra $\Omega$-dependent velocity gradient into the system, contributing to an increase in shear viscosity.

In Figures~\ref{fig-T2} and \ref{fig-T3}, we observe a similar increasing trend of shear viscosity ($\eta$) with temperature ($T$) but with notable differences. In Figure~\ref{fig-T2}, plotted for $\mu = 0.2$ GeV, the curve for $\Omega = 0.1$ GeV interestingly positions itself between the curves for $\Omega = 0.5$ GeV and $1.0$ GeV. The trend completely reverses in Fig.~\ref{fig-T3} as compared to Fig.~\ref{fig-T1}.

These observations reveal a complex interplay between the $\mu$ and $\Omega$. Both factors influence the distribution functions, but $\Omega$ additionally appears in the numerator of the $\eta$ expression, subtly altering its impact. As $\mu$ increases, its influence seems to suppress the effects of $\Omega$. This behavior is constrained by physical limits, particularly causality concerns, which restrict the allowable values of $\Omega$. Excessively large $\Omega$ values could lead to particles escaping the system's confines, violating physical constraints.
These results suggest that while temperature and angular velocity generally enhance each other, leading to potentially higher angular velocities at higher temperatures in collider experiments, the chemical potential exerts a contrasting effect. As $\mu$ increases, the suppression of $\Omega$'s influence becomes more pronounced, indicating a delicate balance between these parameters in the dynamics of rotating fermionic systems like the QGP.
\section{Summary}
\label{sec:summary}
In a medium undergoing finite rotation, transport coefficients are influenced and exhibit a dependency on the medium's angular velocity. Utilizing the Kubo formalism, this study quantifies the shear viscosity in a generalized fermionic system subjected to high angular velocity. The impact of angular velocity is integrated into the calculations through the momentum space propagator of the fermion, obtained via the Fock-Schwinger approach. This influence manifests in the distribution function via Matsubara frequency summation and also appears in the numerator of the shear viscosity expression through trace evaluations. Our findings indicate that shear viscosity rises with increasing temperature and angular velocity, while it diminishes with higher chemical potential. Notably, the chemical potential and its functional equivalent, the angular velocity, exert contrasting effects on the dissipation via shear viscosity in fermionic systems. Research to generalize these findings for arbitrary strengths of angular velocity is ongoing, and those outcomes will be reported in future work.
\section*{Acknowledgements}
We thank Maxim~N.~Chernodub and Amaresh Jaiswal for their encouragement and the useful discussions on this topic and also acknowledge the cooperation received from Abhishek Tiwari and Sumit. S.S. is supported by Institute Post Doctoral Scheme of IIT Roorkee under the grant IITR/Estt-(A)-Rect-Cell-E-5001(130)18490. R.S. acknowledges the support of Polish NAWA Bekker program No. BPN/BEK/2021/1/00342 and Polish NCN Grant No. 2018/30/E/ST2/00432, and thank Francesco Becattini, Derek Teaney, and Kirill Tuchin for the fruitful discussions.
\bibliography{pv_ref}{}
\bibliographystyle{utphys}
\end{document}